%% file: eprint.tex
\documentclass[10pt]{article}
\usepackage{graphicx}

\def\Title#1{\begin{center} {\Large #1 } \end{center}}
\def\Author#1{\begin{center}{ \sc #1} \end{center}}
\def\Address#1{\begin{center}{ \it #1} \end{center}}

\newcommand\pubblock{\rightline{\begin{tabular}{l} Proceedings of the Second Annual LHCP\\ \pubnumber\\
         \pubdate  \end{tabular}}}

\newenvironment{Abstract}{\begin{quotation} \begin{center} 
             \large ABSTRACT \end{center}\bigskip 
      \begin{center}\begin{large}}{\end{large}\end{center} \end{quotation}}

\newenvironment{Presented}{\begin{quotation} \begin{center} 
             PRESENTED AT\end{center}\bigskip 
      \begin{center}\begin{large}}{\end{large}\end{center} \end{quotation}}

\def\Acknowledgements{\bigskip  \bigskip \begin{center} \begin{large}
             \bf ACKNOWLEDGEMENTS \end{large}\end{center}}

\input econfmacros.tex

\usepackage{color}

\newcommand\pubnumber{ CMS-CR-2014/195 }

\newcommand\pubdate{\today}

\def\affiliation{
On behalf of the CMS Collaboration, \\
Department of Physics \\
Brown University, Providence, RI 02912, U.S.A }

\begin{document}

\large
\begin{titlepage}
\pubblock

\vfill
\Title{  Performance of b-tagging algorithms at the CMS experiment with pp collision data at $\sqrt s$=8 TeV  }
\vfill

\Author{ Nitish Dhingra }
\Address{\affiliation}
\vfill
\begin{Abstract}
\begin{center} {The identification of jets originating from b quarks is crucial both for the searches for new physics and for the measurement of standard model processes. The Compact Muon Solenoid (CMS) collaboration at the Large Hadron Collider (LHC) has developed a variety of algorithms to select b-quark jets based on variables such as the impact parameter of charged particle tracks, properties of reconstructed secondary vertices from heavy hadron decays, and the presence or absence of a lepton in the jet, or combinations thereof. Performance measurements of these b-jet identification algorithms are presented, using multijet and $t\overline{t}$ events recorded in proton-proton collision data at $\sqrt s$=8 TeV with the CMS detector during the LHC Run 1.}
\end{center}
\end{Abstract}
\vfill

\begin{Presented}
The Second Annual Conference\\
 on Large Hadron Collider Physics \\
Columbia University, New York, U.S.A \\ 
June 2-7, 2014
\end{Presented}
\vfill
\end{titlepage}
\def\thefootnote{\fnsymbol{footnote}}
\setcounter{footnote}{0}
%

\normalsize 


\section{Introduction}
Jets originating from the hadronization of bottom quarks appear in many important physics processes, such as decays of top quarks, Higgs bosons, and many new particles predicted by supersymmetric models. The identification of b-quark jets is, therefore, a key ingredient in reducing the otherwise overwhelming backgrounds to these channels from the processes involving jets from gluons (g), light flavour quarks (u, d, s), and from c-quark fragmentation. 

This conference report describes various b-jet tagging algorithms developed within the CMS collaboration and the methods used to measure their performance in pp collision data at center-of-mass energy $\sqrt{s}$ = 8 TeV recorded in 2012. For the analyses at $\sqrt{s}$ = 8 TeV, only the Track Counting High Purity (TCHP), Jet Probability (JP), and Combined Secondary Vertex (CSV) algorithms are used, as they are the most efficient (CSV, JP) or matched to a trigger condition with similar algorithms (Combined Secondary Vertex, Track Counting) and loose selection criteria. The description of these algorithms and their performance measurement is presented here, while more details can be found in Ref. \cite{Three:BTV13}. 



\section{Data and Simulation}
Samples of inclusive multijet events were used for the measurement of efficiencies and misidentification probabilities. These samples were collected using single jet triggers with $p_{T}$ thresholds between 40 to 320 GeV/c. For efficiency measurements, dedicated triggers were also used to enrich the data sample with jets from semimuonic b-hadron decays. These triggers required the presence of at least two jets with $p_{T}$ thresholds ranging from 20 to 110 GeV/c, or at least one jet with $p_{T}$ larger than 300 GeV/c. One of these jets was required to include a muon with $p_{T} >$ 5 GeV/c within a cone of $\Delta R$ = 0.4 around the jet axis, where $\Delta R$ is defined as $\sqrt{(\Delta \eta)^{2} + (\Delta \phi)^{2}}$. Here $\phi$ is the azimuthal angle in the (x, y) plane perpendicular to the beam axis and the pseudorapidity is defined as $\eta$ = - ln[tan($\theta$/2)], where $\theta$ is the polar angle relative to the (z) beam axis. All triggers were prescaled except those with the highest-$p_{T}$ thresholds. Data for the analysis of $t\overline{t}$ events were collected with unprescaled single-lepton (e or $\mu$) and double-lepton (ee or e$\mu$ or $\mu\mu$) triggers. For unprescaled triggers, the multijet and $t\overline{t}$ analyses both used datasets with integrated luminosities of 19.8 fb$^{-1}$. Monte Carlo (MC) Simulated samples of multijet events were generated with PYTHIA 6.426 \cite{Four:Pythia} using the Z2 tune \cite{Five:TuneZ2} and $t\overline{t}$ events were simulated using MADGRAPH 5.1.3.30 \cite{Six:MadG}, which was interfaced to PYTHIA for parton showering.





\section{b-tagging algorithms}
A brief description of the b-tagging algorithms used for 2012 data analyses is given below.
\begin{itemize}
\item{{\bf Track Counting High Purity (TCHP):} This algorithm sorts tracks in a jet by decreasing values of their impact parameter (IP) significance (IP/$\sigma_{IP}$, where $\sigma_{IP}$ is the measurement uncertainty on the impact parameter). The discriminator value is defined as the IP significance of the track with the third highest IP significance.}

\item{{\bf Jet Probability (JP):} As a natural extension of the Track Counting algorithm, the JP algorithm combines the IP information from all selected tracks in the jet. The jet is assigned a likelihood that all associated tracks come from the primary vertex (PV). The probability distribution of individual tracks is computed using the IP significance of tracks with negative IP values, which are mostly produced in light-parton jets; this calibration is performed independently in data and simulation.}

\item{{\bf Combined Secondary Vertex (CSV):} This algorithm combines secondary vertex and displaced track information to build a likelihood-based discriminator to distinguish between jets from b quarks and those from charm or light quarks and gluons.}
\end{itemize}

\begin{figure}[htb]
\centering
\includegraphics[height=1.7in]{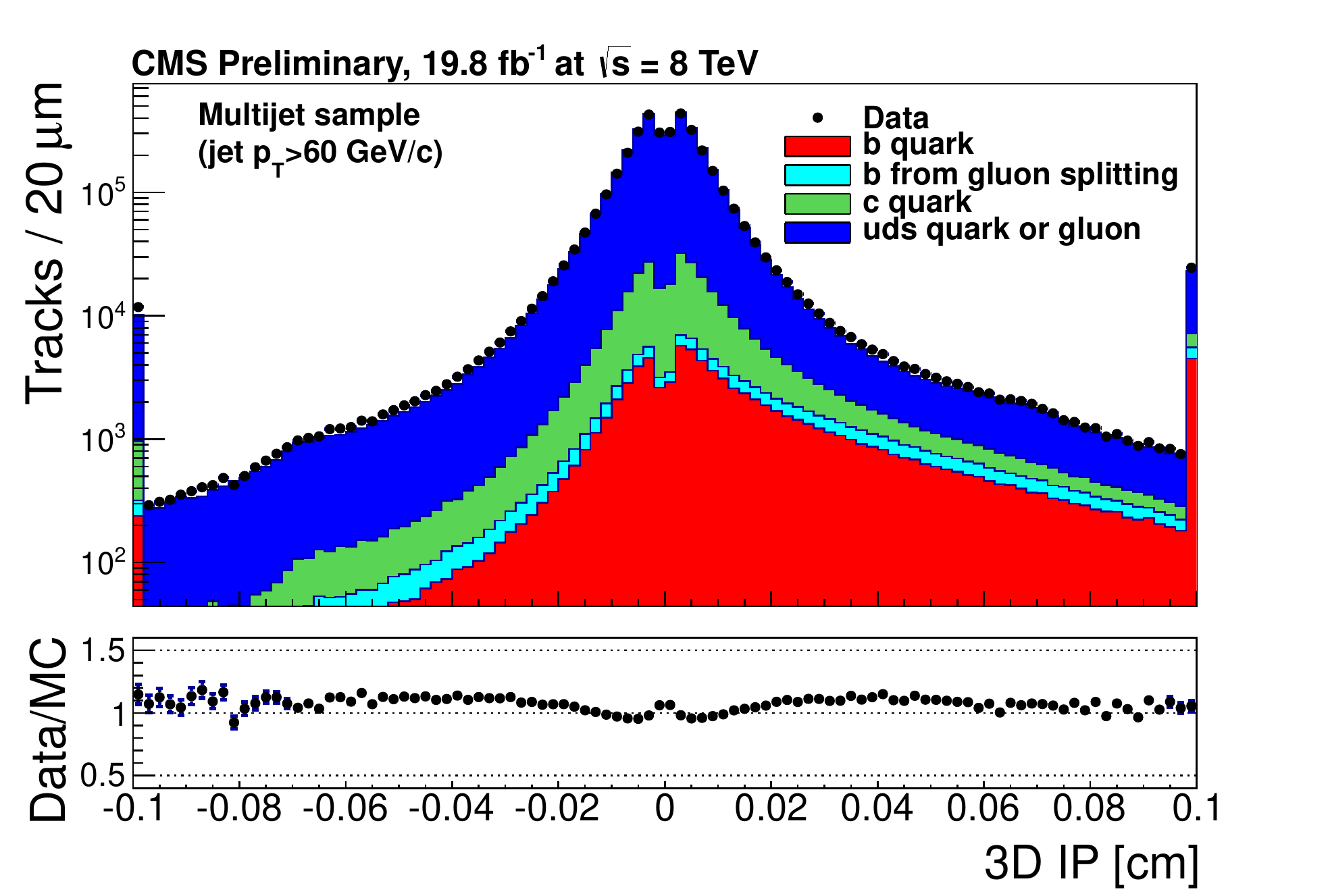}
\includegraphics[height=1.7in]{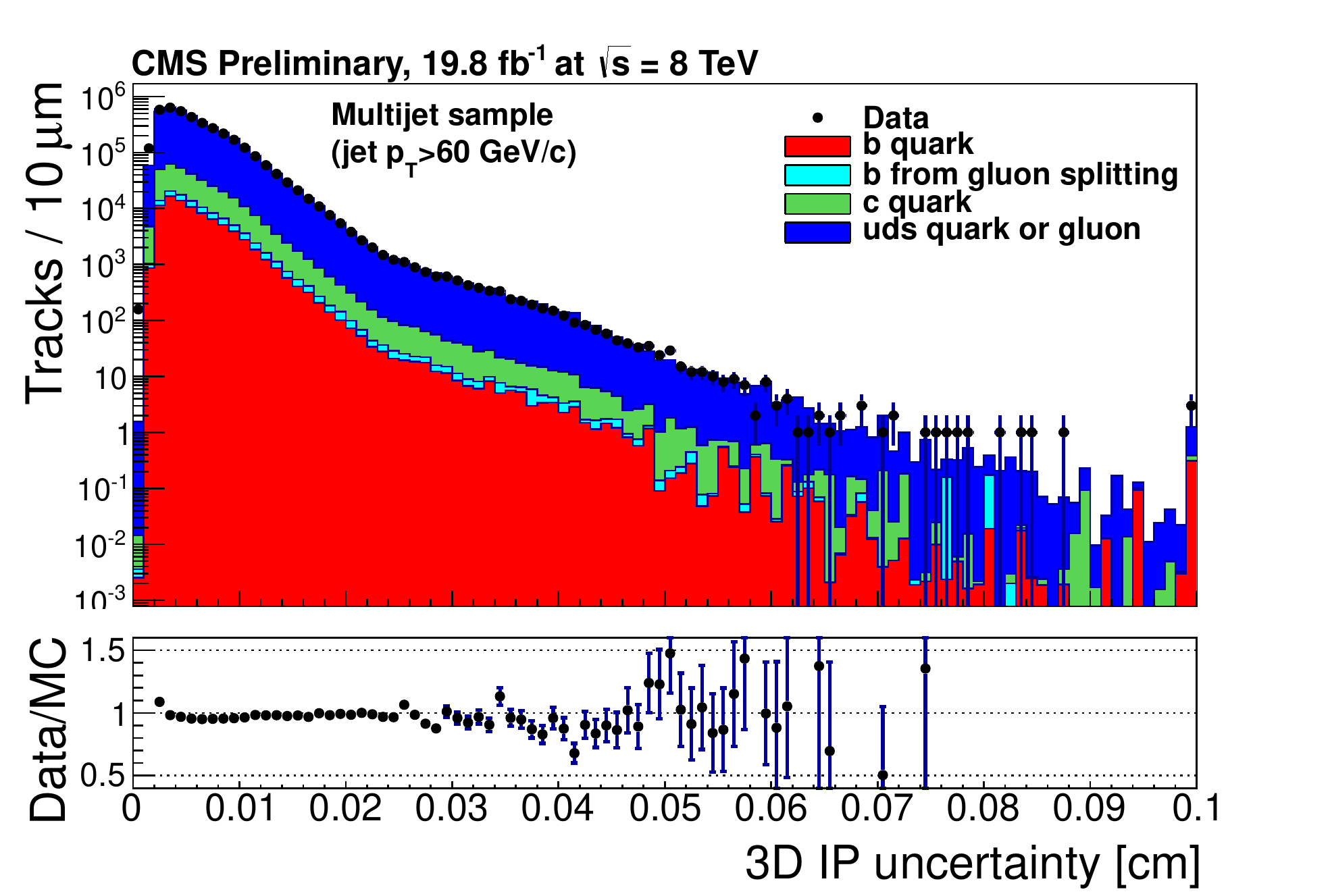} \\
 (a) \hspace{2.4in}        (b) \\
\includegraphics[height=1.7in]{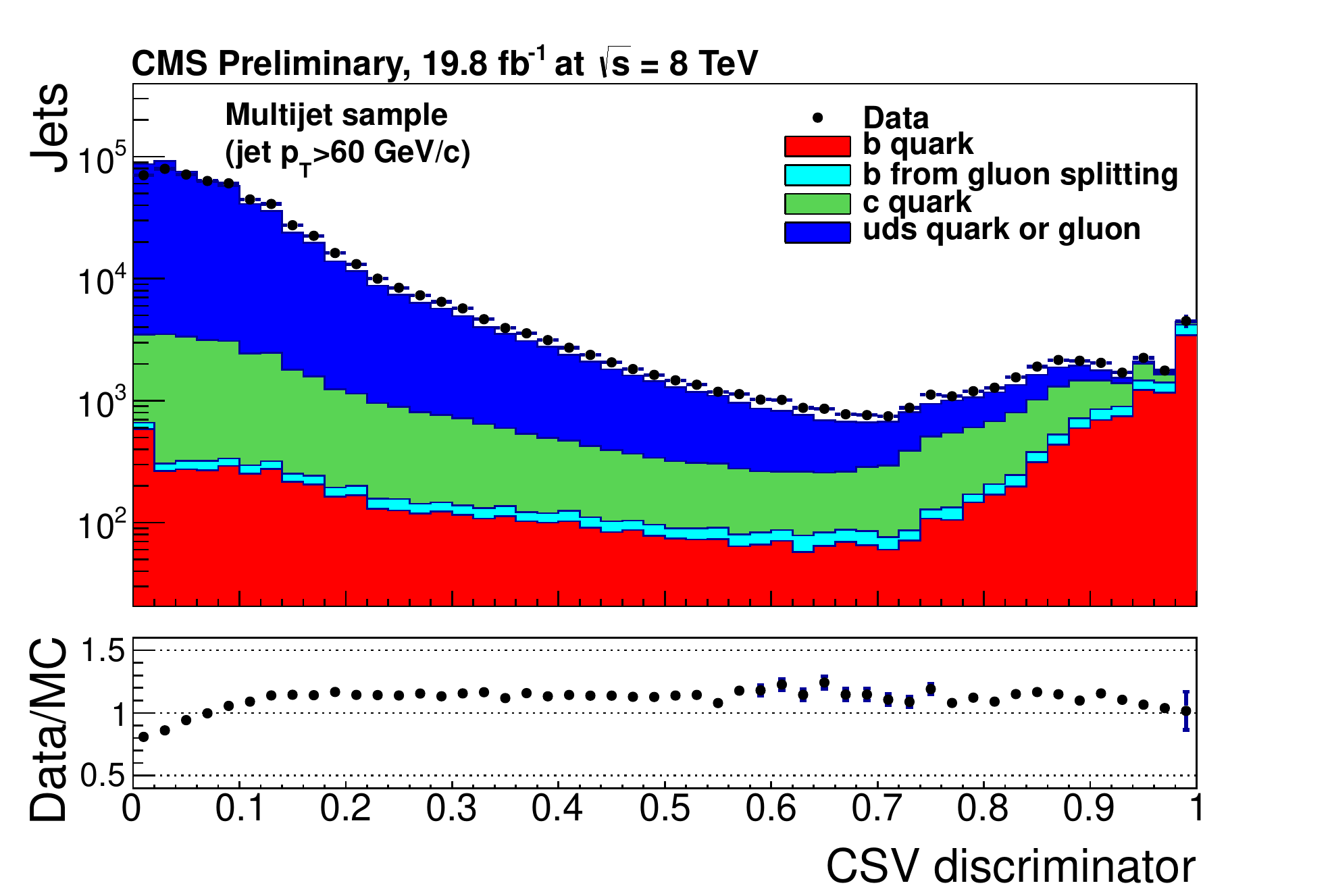}
\includegraphics[height=1.7in]{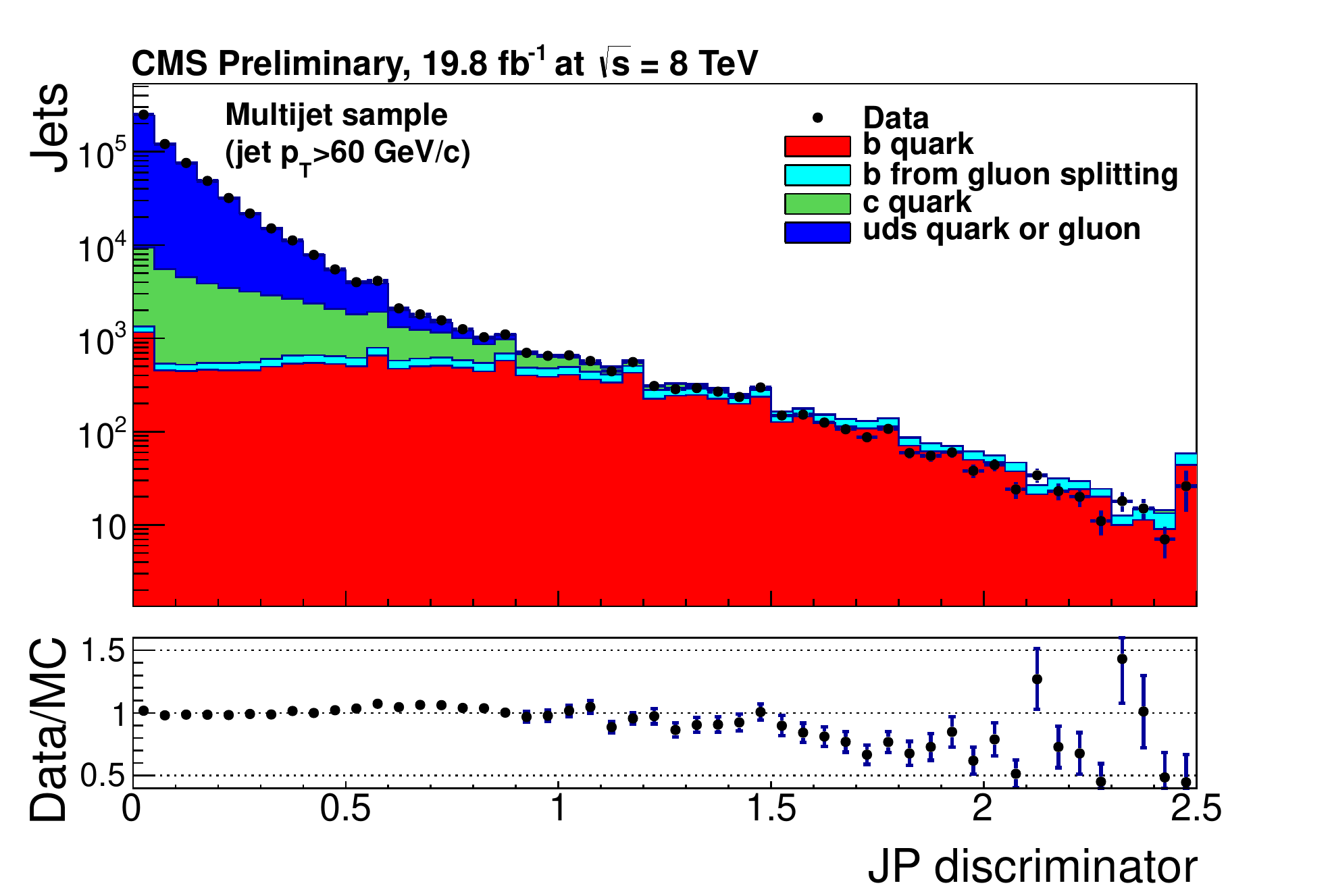} \\
 (c)  \hspace{2.4in}        (d) \\
\caption{ Distribution of (a) the 3D impact parameter, (b) the uncertainty on the 3D impact parameter for all selected tracks, (c) CSV discriminator and (d) JP discriminator in the inclusive multijet data sample. The simulation distributions for light-parton jets (blue histogram), c jets (green histogram) and b jets (red histogram) are also overlayed.}
\label{fig:figure1}
\end{figure}

The minimum thresholds on these discriminators define loose (“L”), medium (“M”), and tight (“T”) operating points with a misidentification probability for light-parton jets of close to 10$\%$, 1$\%$, and 0.1$\%$, respectively, with an average jet $p_{T}$ of about 80 GeV/c. The first letter of the used operating point is usually appended to the acronym of the algorithm to fully define the tagging criterion. For instance, JPL indicates the loose operating point for the JP tagger. The distributions of the IP and its uncertainty as well as the distributions of the JP and CSV discriminators are shown in Figure \ref{fig:figure1} for inclusive multijet event sample. All these distributions exhibit a good agreement between data and simulation, with deviations generally within 20$\%$, while larger discrepancies are observed only in the tails of the distributions. 

\section{Misidentification probability measurement}
\begin{figure}[htb]
\centering
\includegraphics[height=2.2in]{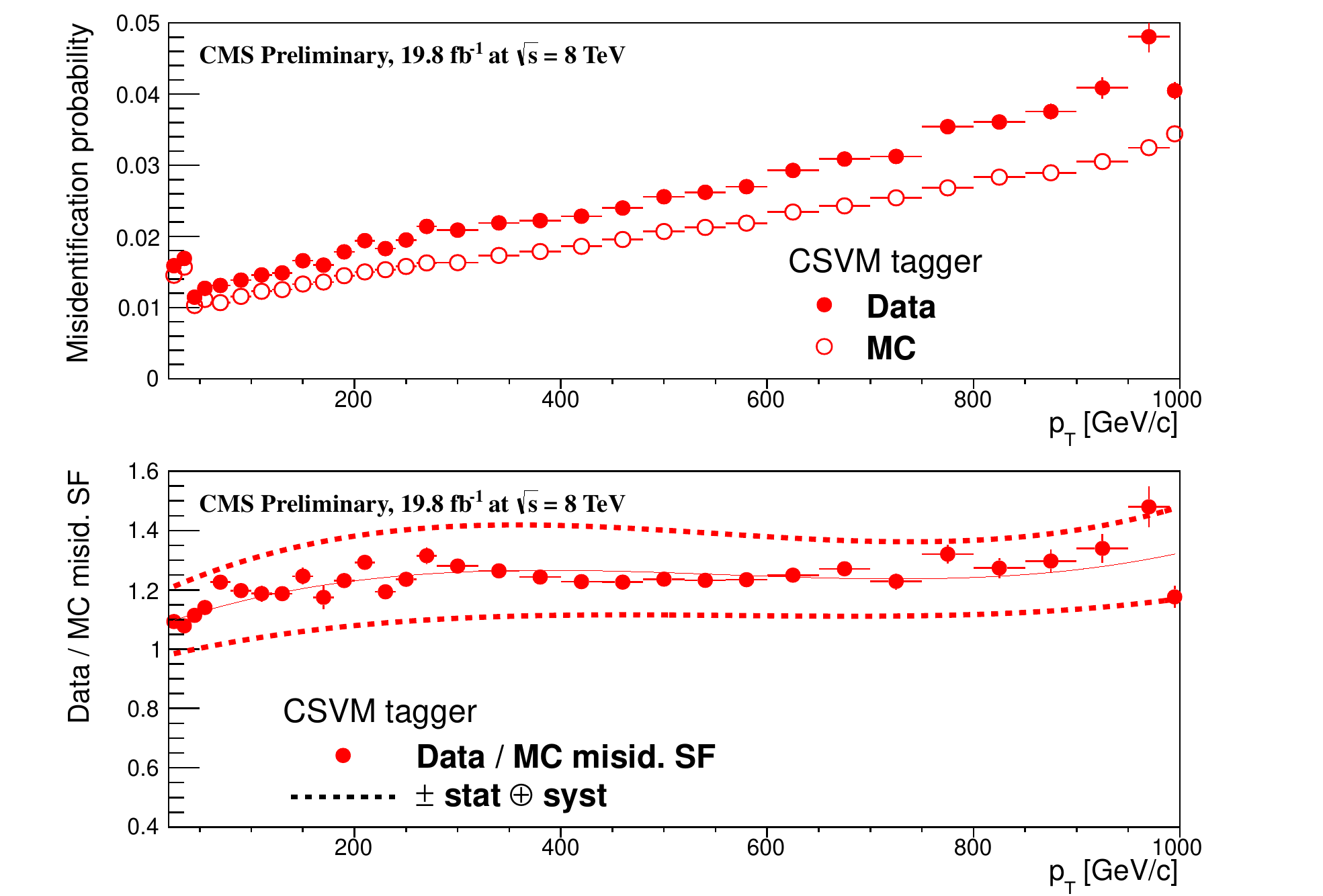}
\caption{ For the CSVM tagging criteria: (top) misidentification probability in data and simulation; (bottom) scale factor for the misidentification probability. The last $p_{T}$ bin in each plot includes all jets with $p_{T} > 1000$ GeV/c. The solid curve is the result of a polynomial fit to the data points. The dashed curve represents the overall statistical and systematic uncertainties on the measurements. }
\label{fig:figure2}
\end{figure}

The probability of light-flavour quark and gluon jets being misidentified as b-jets, $\epsilon^{misid}$, is evaluated with negative tagging algorithms. The negative taggers are identical to the default algorithms, except that they are used only on tracks with negative IP values or on secondary vertices with negative decay lengths. To first order, the discriminator values for negative and positive taggers are expected to be symmetric for light-parton jets, since they are due only to resolution effects in track reconstruction. We can therefore derive the misidentification probability from the rate, $\epsilon^{-}$, of negative-tagged jets in inclusive jet data. A correction factor, $R_{light}$, is evaluated from the simulation in order to correct for second-order asymmetries in the negative
and positive tag rates of light-flavour quark and gluon jets, and for the heavy flavour contribution to the negative tags: $\epsilon^{misid}_{data} = \epsilon^{-}_{data}\cdot R_{light}$ where $R_{light} = \epsilon^{misid}_{MC}/\epsilon^{-}_{MC}$. Using different jet-trigger $p_{T}$ thresholds, the misidentification probability can be computed in a wide jet $p_{T}$ range from 20 to 1000 GeV/c, as shown in Figure \ref{fig:figure2} for the CSVM criterion. The scale factor $SF_{light}= \epsilon^{misid}_{data}/\epsilon^{misid}_{MC}$ is calculated as the ratio of misidentification probability in data and simulation (MC).

\section{b-tagging efficiency measurement}
The b-tagging efficiency is measured from data using a control sample of jets enriched in heavy flavour content. 
The control sample is obtained by selecting events with a soft-muon inside the jet within a cone-size $\Delta$R $<$0.4 around jet-axis and another tagged-jet (``away tag'') to reduce the light-flavour background. To determine the b-tagging efficiency, methods are used that are based on kinematic properties of the muon-jets, {\it e.g.} $p_{T}^{\mu}$ relative to jet axis (PtRel), the IP of the muon (IP3D) and Lifetime Tagging (LT). $t\overline{t}$ events can also be used but due to event kinematics they are less suited for an efficiency measurement in the high jet $p_{T}$ range. The muon-jets are separated into tagged and untagged subsamples by a discriminator working point whose efficiency is to be measured. For the two subsamples separately, the spectra of muon-jets $p_{T}^{rel}$ or IP3D are fitted using templates of b, c, and udsg jets derived from simulation or inclusive jet data. The fraction of b-jets in the two subsamples, $f^{tag}_{b}$ and $f^{untag}_{b}$, are then calculated and the efficiency is determined as: 
$\epsilon^{tag}_{b} = \frac{f^{tag}_{b}\cdot N^{tag}_{data}}{f^{tag}_{b}\cdot N^{tag}_{data} + f^{untag}_{b}\cdot N^{untag}_{data}}$.
\vspace{0.3cm}

The b-tagging efficiency and data/MC b-tagging efficiency scale factor for the CSVM criterion as a function of jet $p_{T}$ and jet $|\eta|$ is shown in Figure \ref{fig:figure3}. An overview of the individual and combined measurement of the b-tagging efficiency scale factor for the CSVM criterion obtained using various methods is shown in the Figure \ref{fig:figure4}. In the same figure, the parameterisations of the combined scale factor of the form SF$_{b}$($p_{T}$) = $\alpha$(1 + $\beta p_{T}$)/(1 + $\gamma p_{T}$) is also shown. The details of System8, J/$\Psi$, and other methods can be found in Ref. {\cite{Seven:BTV12}.

\begin{figure}[htb]
\centering
\includegraphics[height=2.05in]{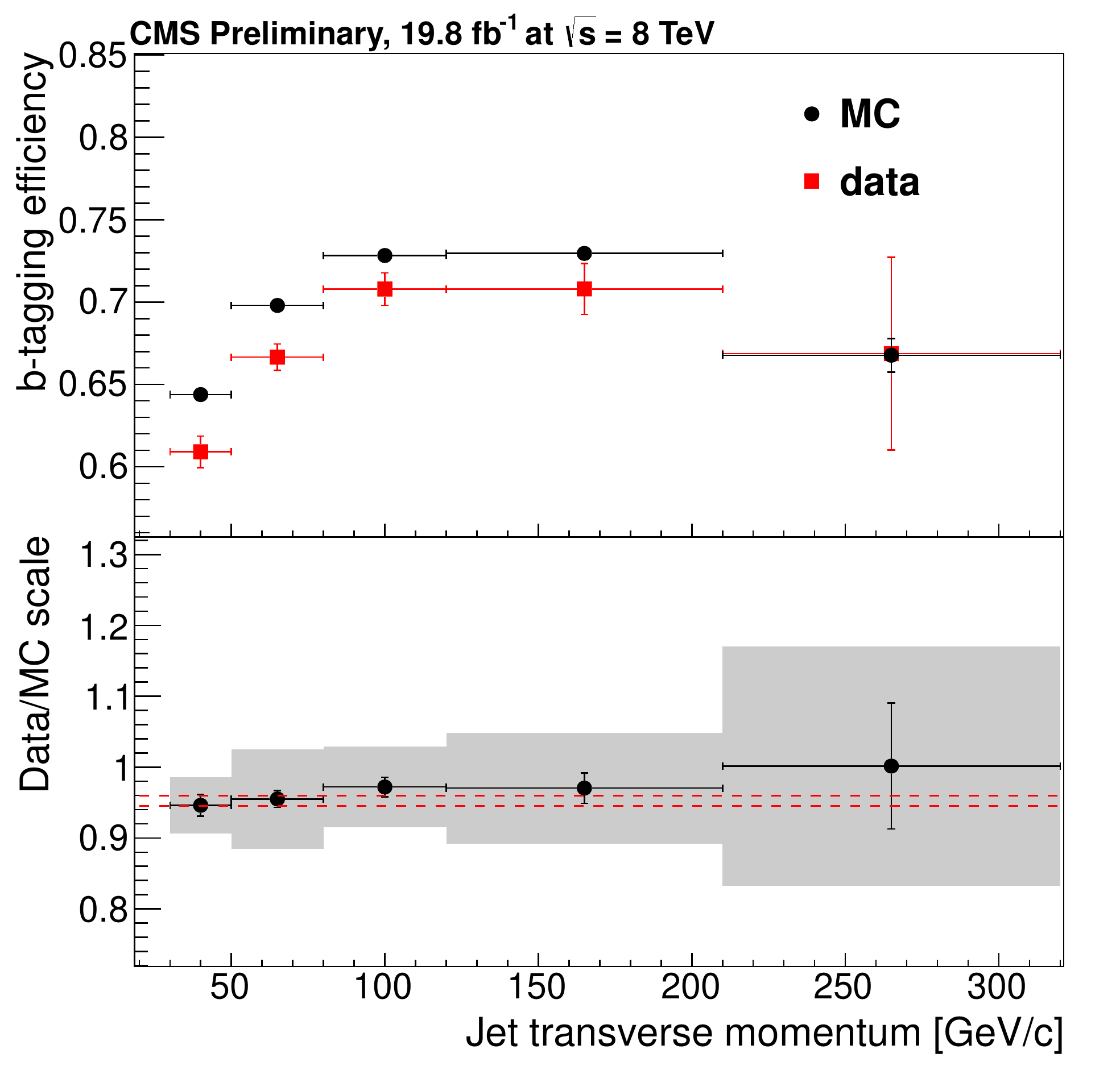}
\includegraphics[height=2.05in]{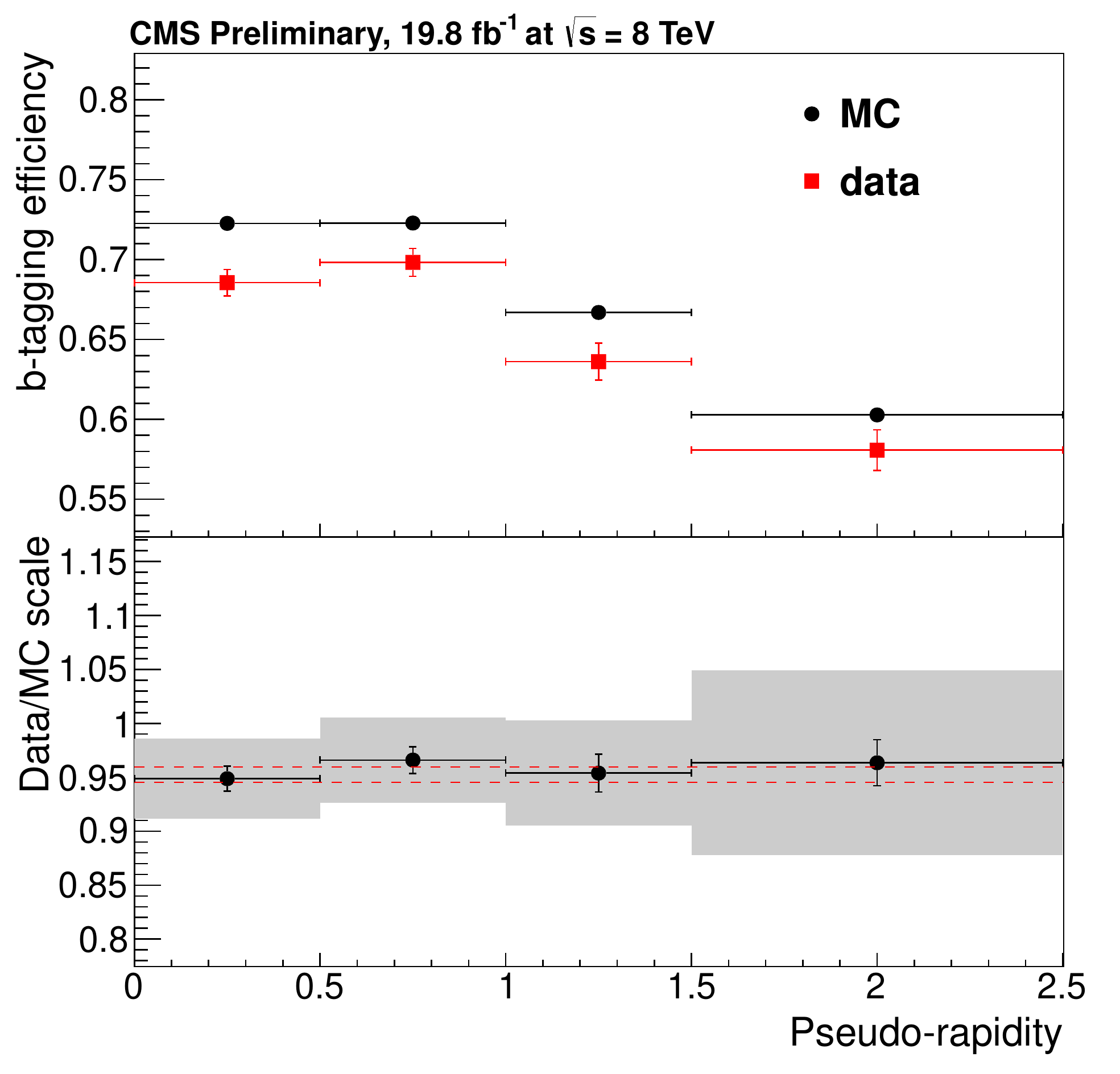}
\caption{ In $t\overline{t}$ dilepton events with LTtop method for CSVM criterion: b-tagging efficiencies and data/MC b-tagging efficiency scale factor SF$_{b}$ as a function of (left) jet $p_{T}$ and (right) jet $|\eta|$. The grey filled area represents the total statistical and systematic uncertainties, whereas the dotted lines are the average $SF_{b}$ values within statistical uncertainties.}
\label{fig:figure3}
\end{figure}

\begin{figure}[htb]
\centering
\includegraphics[height=2.3in, width=3.0in]{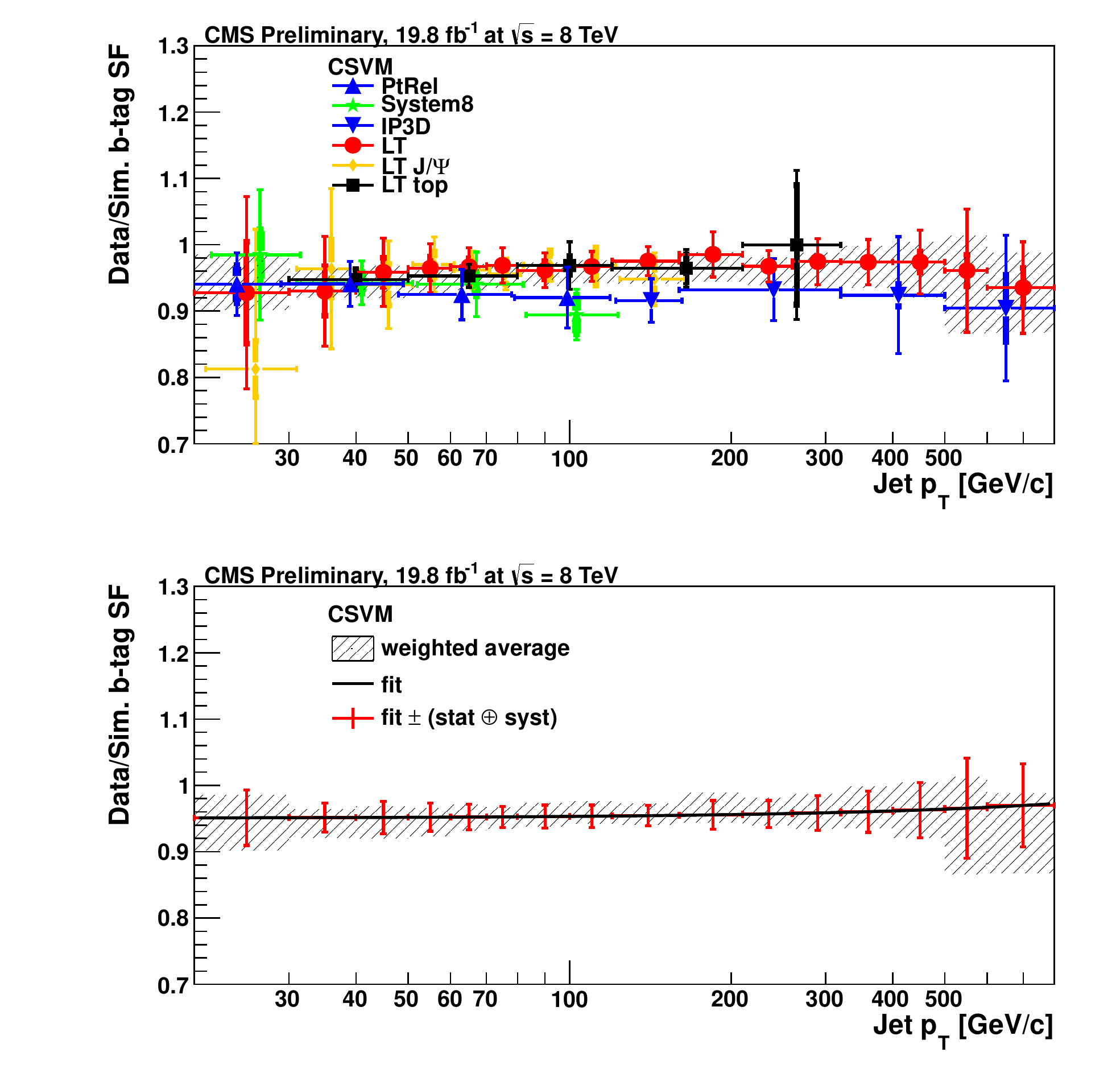}
\caption{ Individual (top) and combined (bottom) measurements of the data/MC scale factor $SF_{b}$ of the b-tagging efficiencies for the CSVM tagging criterion. The individual measurements are obtained using the muon $p_{T}^{rel}$, System8, muon IP3D, lifetime tagger (``LT'') and J/$\Psi$ methods.}
\label{fig:figure4}
\end{figure}



\section{Conclusions}
Several methods have been developed by the CMS collaboration to measure the b-quark jet identification efficiency and misidentification probability. The results obtained by different methods are found to be consistent with each other. Overall, the b-tagging performance measurements at $\sqrt s$ = 8 TeV are in good agreement with those observed at 7 TeV. 

\Acknowledgements
I am extremely grateful to Dr. Alexander Schmidt, Dr. Luca Scodellaro, Dr. Petra Van Mulders, and the members of BTV group at CMS for the fruitful discussions and suggestions.

\end{document}

%% file: econfmacros.tex
\def\beq{\begin{equation}}
\def\eeq#1{\label{#1}\end{equation}}
\def\eeqn{\end{equation}}

\def\beqa{\begin{eqnarray}}
\def\eeqa#1{\label{#1}\end{eqnarray}}
\def\eeqan{\end{eqnarray}}

\let\bar=\overbar

\def\Dslash{\not{\hbox{\kern-4pt $D$}}}
\def\dslash{\not{\hbox{\kern-2pt $\del$}}}

\def\msb{{\bar{\ssstyle M \kern -1pt S}}}

